\documentclass[conference]{IEEEtran}
\IEEEoverridecommandlockouts
\usepackage[table,xcdraw]{xcolor}
\usepackage{colortbl}

\usepackage{amsmath,amssymb,amsfonts}
\usepackage{algorithmic}
\usepackage{graphicx}
\usepackage{authblk}
\usepackage{subfigure}
\usepackage{subfig}
\usepackage{caption}
\usepackage{array}

\usepackage[numbers]{natbib}

 \makeatletter

\newcommand{\Rmnum}[1]{\expandafter\@slowromancap\romannumeral #1@}
\makeatother
\usepackage{textcomp}
\usepackage{xcolor}
\def\BibTeX{{\rm B\kern-.05em{\sc i\kern-.025em b}\kern-.08em
    T\kern-.1667em\lower.7ex\hbox{E}\kern-.125emX}}
\begin{document}

\title{A Multi-Scale Spatial-Temporal Network for Wireless Video Transmission}

\author[1]{Xinyi~Zhou}
\author[*1]{Danlan~Huang\thanks{*This is corresponding author. huangdl@bupt.edu.cn.}}
\author[1]{Zhixin~Qi}
\author[1]{Liang~Zhang}
\author[1]{Ting~Jiang}

\affil[1]{Beijing University of Posts and Telecommunications, Beijing, China}

\maketitle

\begin{abstract}
Deep joint source-channel coding (DeepJSCC) has shown promise in wireless transmission of text, speech, and images within the realm of semantic communication. However, wireless video transmission presents greater challenges due to the difficulty of extracting and compactly representing both spatial and temporal features, as well as its significant bandwidth and computational resource requirements. In response, we propose a novel video DeepJSCC (VDJSCC) approach to enable end-to-end video transmission over a wireless channel. Our approach involves the design of a multi-scale vision Transformer encoder and decoder to effectively capture spatial-temporal representations over long-term frames. Additionally, we propose a dynamic token selection module to mask less semantically important tokens from spatial or temporal dimensions, allowing for content-adaptive variable-length video coding by adjusting the token keep ratio. Experimental results demonstrate the effectiveness of our VDJSCC approach compared to digital schemes that use separate source and channel codes, as well as other DeepJSCC schemes, in terms of reconstruction quality and bandwidth reduction.

\end{abstract}


\section{Introduction}
With the rapid emergence of video traffic such as video conference and virtual reality, 
wireless video transmission technology has gained widespread attention to tackle large volume data. Digital wireless video transmission scheme is based on separate source and channel coding. The primary source codes such as industry standard video codecs H.264/H.265 attempt to eliminate redundant information, while the channel codes such as low density parity check (LDPC) adopts redundant bits to overcome distortions in imperfect wireless channel. However, the separate coding scheme is sub-optimal for video transmission since it suffers from \textit{cliff-effect} and is hard to meet the low latency requirement of video applications.


\par To address the aforementioned issues, joint source-channel coding (JSCC) has been proposed to achieve system-level optimality in the coding process \cite{6408177}. Owing to the rapid advancement of deep learning (DL) in recent years, an increasing number of DL models are being used in JSCC systems to enhance compact feature representation and noise resilient capabilities. Deep joint source-channel coding (DeepJSCC) scheme has demonstrated significant success in various wireless data transmission tasks, including image \cite{9791398}, text \cite{8461983}, and speech \cite{9450827}. Nevertheless, there is still limited research on video wireless transmission task. The core challenge is how to capture the dynamic temporal features across frames, instead of merely considering the static spatial feature within a frame. 

In order to facilitate the wireless transmission of video, the incorporation of the image DeepJSCC method \cite{9791398,10094735} can be used to individually process each frame. However, this approach overlooks the temporal correlations among sequential frames and applies an identical compression ratio to each frame, resulting in unnecessary redundancy. Alternatively, previous research \cite{9837870,10112629} has focused on separately encoding key frames and residual motion information in order to account for temporal correlations. For instance, MGCNet \cite{10112629} utilizes the extraction and aggregation of context features from per-frame, short-term, and long-term granularities. Nevertheless, these approaches often rely on the calculation of motion information through the use of optical flow algorithms, which are burdened by heavy computational complexity and limit the methods' practical generalization.

Recently, the Vision Transformer (ViT) \cite{dosovitskiy2020image} has demonstrated promise for image analysis tasks. Additionally, the Video Swin Transformer \cite{9878941} has expanded the applicability of local attention computation from spatial to spatial-temporal domains by incorporating 3D patch and shifted window techniques. The robust feature representation abilities of the Video Swin Transformer \cite{9878941} enable effective management of correlations between temporal and spatial dimensions. Nevertheless, the transmission of 3D tokens presents a considerable bandwidth challenge for wireless communication.

\begin{figure*}[htb]
\centering
{\includegraphics[width=1.0\textwidth]{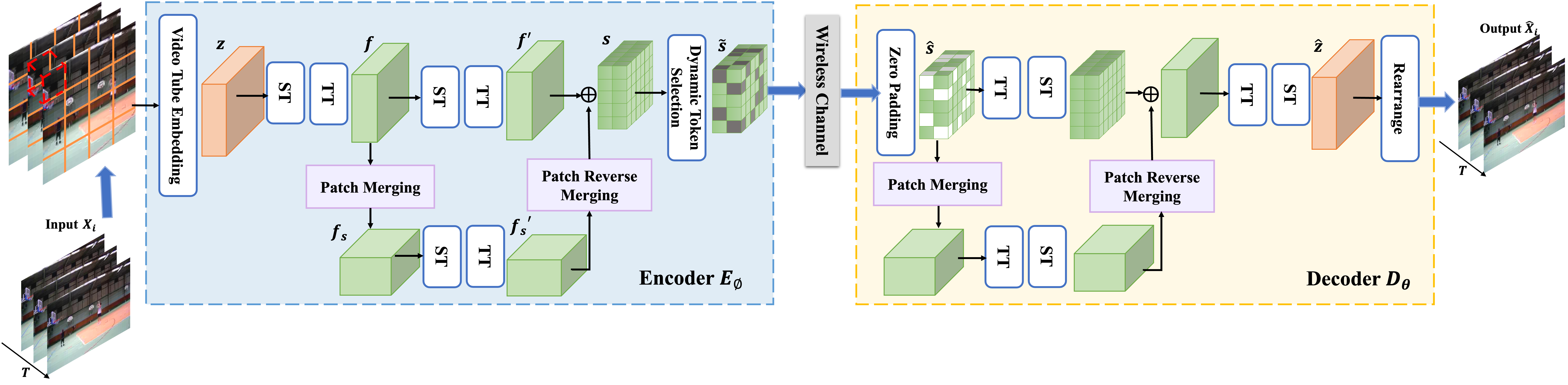}}
\caption{The architecture of the proposed VDJSCC scheme, where ST/TT refer to spatial/temporal Transformer, respectively.}
\label{fig1}
\end{figure*}

To address the aforementioned challenges, this paper presents a novel video DeepJSCC scheme (VDJSCC) that incorporate spatial-temporal attention to encode and transmit video frames. We innovatively adopt Vision Transformer to capture semantic information in video transmission. The contributions can be summarized as follows:


\begin{itemize}

\par \item  We introduce a novel video DeepJSCC scheme, referred to as VDJSCC, which utilizes the multi-scale vision Transformer for encoding spatial-temporal features. Within the VDJSCC model, we incorporate a spatial-temporal separation attention mechanism to capture comprehensive semantic information.
\par \item  In order to optimize computational efficiency and reduce bandwidth burden, we have developed a dynamic token selection module, which identifies and masks tokens with lower semantic importance. This module also adjusts the encoding length based on the token keep ratio, enabling content-adaptive variable-length coding.
\par \item   Our experimental findings demonstrate that the proposed VDJSCC scheme outperforms traditional methods, such as H.264 combined with LDPC and digital modulation schemes, across established performance metrics including peak signal-to-noise ratio (PSNR) and multiscale structural similarity index (MS-SSIM). Furthermore, VDJSCC offers significant advantages in terms of transmission resource savings.
\end{itemize}


\section{The proposed VDJSCC scheme}
In this section, we will first present the overall structure of the proposed VDJSCC scheme. After that, the detailed content of each module will be introduced separately.

\subsection{The Overall Architecture of VDJSCC}
The architecture of the proposed video wireless transmission scheme, noted as VDJSCC, is illustrated in Fig. \ref{fig1}. We consider the wireless transmission of videos over the additive white Gaussian noise (AWGN) channel. Denoting $T$ as the number of frames, $C$ as the number of color channels, $H$ and $W$ as the height and the width of the frame. The video sequences can be represented as $\mathbf{X}_i=\left \{ \mathbf{x}_1, \mathbf{x}_2,...,\mathbf{x}_T  \right \} $, where $\mathbf{X}_i\in\mathbb{R} ^{T\times C\times H\times W} $.
The proposed VDJSCC model includes a pair of trainable encoder $  \mathbf{E_{\phi }} $ and decoder $\mathbf{D_{\theta }}$  and a non-trainable physical channel, where $\phi$ and $\theta$ are the parameters for the encoder and decoder, respectively.  

 Inspired by the patch embedding in ViT \cite{dosovitskiy2020image}, video vision Transformer (ViViT) \cite{9710415} first used the tubelet embedding to extract non-overlapping, spatial-temporal tubes from the input video sequence. The video sequence $\mathbf{X}_i$ is initially split into video tubes. Subsequently, these tubes are then flattened and converted to tokens $\mathbf{z}\in\mathbb{R} ^{n_{t}\times{n_{h}\times n_{w}\times{K} }}$ through a trainable linear projection, where $K$ is the hidden dimension. For a video tube of dimension $t\times h\times w$, $n_{t}=\left \lfloor \frac{T}{t}  \right \rfloor $, $n_{h}=\left \lfloor \frac{H}{h}  \right \rfloor  $, $n_{w}=\left \lfloor \frac{W}{w}  \right \rfloor  $, tokens are extracted from the temporal, height, and width dimensions respectively. The tokens $\mathbf{z}$ is processed by both spatial Transformer (ST) and temporal Transformer (TT) in sequence, resulting in the generation of feature $\mathbf{f}\in\mathbb{R} ^{n_{t}\times(n_{h}\times n_{w})\times{K} }$. This approach allows for the effective capture of multi-scale features aimed at enhancing the representation of details. The down-scaled feature $\mathbf{f}_s\in\mathbb{R} ^{n_{t}\times(\frac{n_{h}}{2}\times \frac{n_{w}}{2})\times{2K}}$ is obtained through a patch merging process, followed by further processing with ST and TT. Ultimately, the outputs from these separate branches are aggregated through averaging, resulting in the generation of feature $\mathbf{s}\in\mathbb{R} ^{n_{t}\times(n_{h}\times n_{w})\times{K} }$. 

In order to reduce the amount of tokens to be transmitted, we design a dynamic token selection module, which mask tokens with less semantic importance at a certain token keep ratio $\gamma$. We denote the pruned tokens that required to be transmitted as $\mathbf{\tilde{s}}$, and the mask matrix is represented as $\Omega\in \{0,1\}^{M}$, where $M$ is the amount of video tube. Additionally, the power normalization operation enables $\mathbf{\tilde{s}}$ to satisfy the average power constraint before transmitting into the channel. The Encoder can be formulated as:

\begin{equation}
    (\mathbf{\tilde{s}}, \Omega)=\mathbf{E_{\phi }}(\mathbf{X}_i,\gamma).
\end{equation}



The wireless channel can be formulated as $\mathbf{\hat{s}}=h\mathbf{\tilde{s}}+\mathbf{n}$, where $h$ denotes the channel gain coefficient. In this formula, $\mathbf{n}\in \mathcal{CN}\sim(0,\sigma ^{2}\mathrm {I})$ denotes independent identically distributed (i.i.d) AWGN samples with power $\sigma ^{2}$. Assuming that the dimension of the original data is $N=T\times C\times H\times W$, and $R=(\gamma\times  K)/N$ is channel bandwidth ratio (CBR).

\par At the receiver, the VDJSCC Decoder $\mathbf{D_{\theta}}$ expands the received tokens $\mathbf{\hat{s}}$ with zero elements to maintain dimensional consistency. Finally, the TT and ST are utilized to reconstruct the input video sequence $\hat{\mathbf{X}}_i\in  \mathbb{R}^{T\times C\times H\times W } $, which can be formulated as:
\begin{equation}
    \hat{\mathbf{X}}_i=\mathbf{D_{\theta}}(\mathbf{\hat{s}}, \Omega, \gamma).
\end{equation}

\par The VDJSCC model is trained in an end-to-end manner. We optimize the model by the mean square error (MSE) between $\mathbf{X}_i$ and $\hat{\mathbf{X}}_i$. The training loss function is formulated as:
\begin{equation}
     \mathcal{L}_{\phi,\theta,\gamma} =\left \| \mathbf{X}_i-\hat{\mathbf{X}}_i \right \| _{2}^{2}. 
\end{equation}

\noindent The optimal model parameters can be obtained by:
\begin{equation}
    \begin{aligned}
   \underset{\phi,\theta,\gamma}{\mathrm{argmin}} \quad & \mathcal{L}_{\phi,\theta,\gamma}\\
    \text{s.t.} \quad\quad & r(\mathbf{\tilde{s}} )+r(\Omega )\le \Gamma, 
\end{aligned}
\end{equation}

\noindent where $ r(\mathbf{\tilde{s}} )$ and $r(\Omega )$ denote the coding rate of the token and the mask matrix. $\Gamma $ means a certain transmission coding rate.

\par In the video reconstruction task, accounting for temporal correlations is essential to minimize redundant information. The proposed VDJSCC scheme effectively captures multi-scale spatial-temporal representations, encompassing temporal information over long-term frames and spatial information within individual frames. While multi-scale feature extraction demands more computational resources, this paper focuses on two specific feature scales. Additionally, the dynamic token selection block generates an adaptive mask matrix $\Omega$ based on the video content to discard less significant tokens, thereby conserving transmission resources. By employing these approaches, VDJSCC can effectively reduce the information redundancy and save the transmission resources.

\subsection{Spatial-Temporal Transformer Module}


\begin{figure}[t]
\centering
{\includegraphics[width=0.45\textwidth]{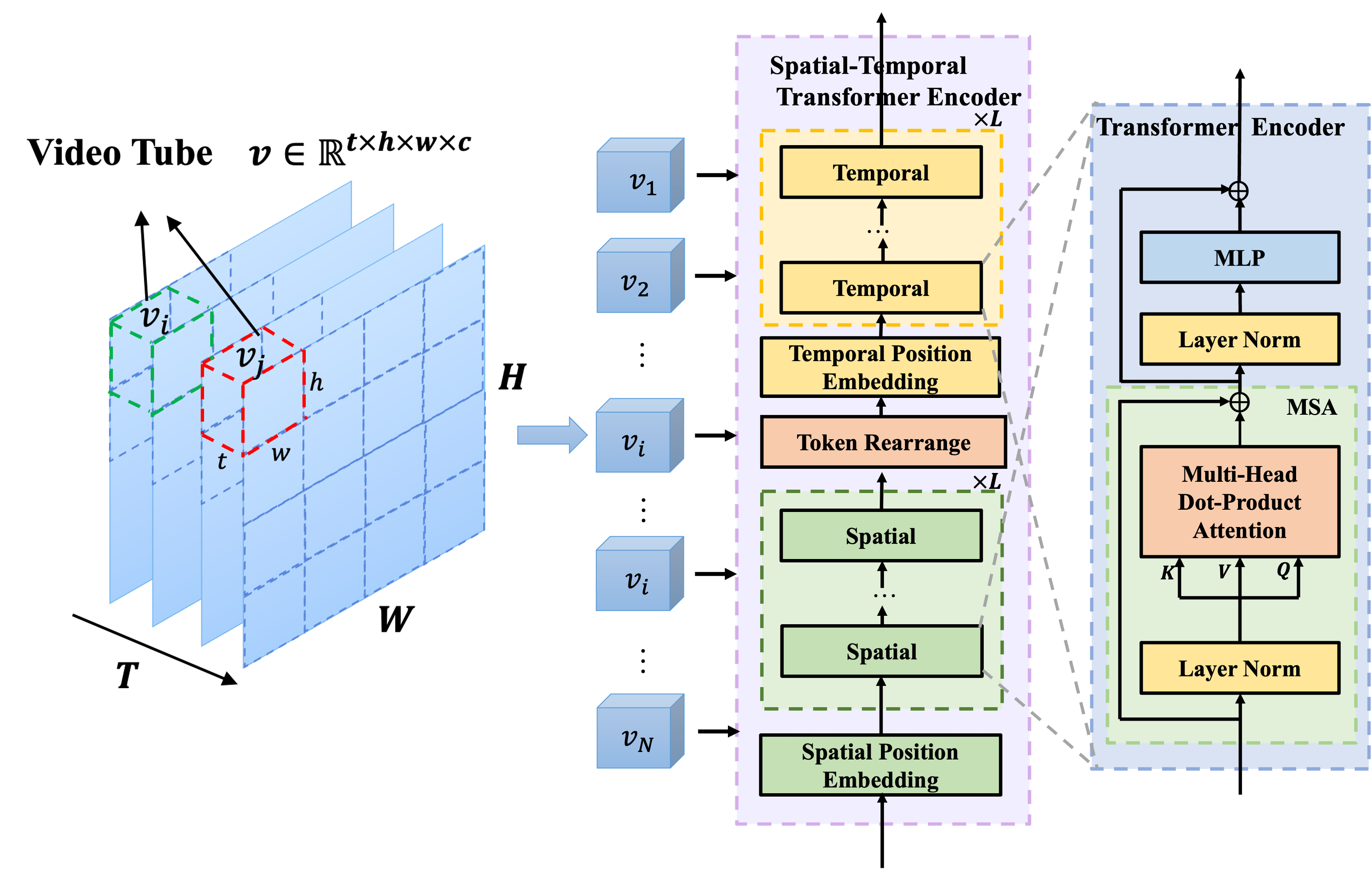}}
\caption{The structure of spatial-temporal Transformer encoder.}
\label{fig}
\end{figure}

 \noindent\textbf{1) Spatial-Temporal Transformer Encoder and Decoder}

 \par As is shown in Fig. 2, the spatial-temporal Transformer encoder consists of two separate Transformer encoders, the ST encoder and the TT encoder. Firstly, spatial position embedding is added to retain spatial positional information. After $L$-layer ST, we rearrange the tokens from $n_{t}\times({n_{h}\times n_{w})\times{K} }$ to $({n_{h}\times n_{w})\times n_{t}\times{K} }$ to make the model pay more attention to temporal connection. For the TT, the temporal position embedding is also added to obtain temporal positional information. Finally, the tokens are processed through $L$-layer TT to obtain the final output. 
 \par In the spatial-temporal Transformer encoder, each layer $l$ contains the multi-head self-attention (MSA), layer normalization (LN), and the multilayer perceptron (MLP) blocks. Taking feature $\mathbf{f}$ as an example, this process can be written as follows:
 \begin{equation}
     \mathbf {y}   ^{ l }=\mathrm {MSA}   \left (\mathrm {LN}   \left(\mathbf {f}  ^{l} \right)\right)+ \mathbf {f}   ^{l},
 \end{equation}
\begin{equation}
     \mathbf {f}   ^{ l+1 }=\mathrm {MLP}  \left  (\mathrm {LN}   \left (\mathbf {y}   ^{l} \right )\right )+ \mathbf {y}   ^{l}.
\end{equation}

\noindent In the MSA, we perform attention operation for each head as follows:
\begin{equation}
    \mathrm {Attention}(\mathrm {Q}, \mathrm {K},\mathrm {V})=\mathrm {Softmax}\left (\frac{\mathrm {Q}\mathrm {K}^{\top } }{\sqrt{d_{k} } }\right )\mathrm {V}.
 \end{equation}

In the spatial-temporal Transformer decoder, we incorporate both spatial and temporal position embedding to capture positional information. Initially, the tokens $\mathbf{\hat{s}}$ and the corresponding downscaled features are processed by TT, after which they are rearranged and fed into ST. Subsequently, the two multi-scale branches are merged and processed by another TT and ST, resulting in the generation of reconstructed tokens $\mathbf{\hat{z}}$.

\noindent\textbf{2) Multi-Scale Transformer Encoder and Decoder}


\par In order to enhance the feature representation capability, we down sample the video frames to achieve multi-scale features. Inspired by the downsampling method in Swin Transformer \cite{9710580}, we use patch merging to downsample the video frames. The patch merging layer concatenates the features of $2\times2$ neighboring patches, and applies a linear layer on the $4K$-dimensional concatenated features. This patch merging process reduces the number of tokens by a multiple of $2\times2=4$ ($2\times$ downsampling of resolution), and the output dimension is set to $2K$. Additionally, the patch reverse merging process reverts the downsampled tokens to their original dimensions, thereby preserving the dimensional consistency between the input and output tokens.
At the receiver, we continue to use patch merging to downsample the tokens $\mathbf{\hat{s}}$ and employ patch reverse merging to restore the original dimensions.

\begin{figure}[t]
\centering
{\includegraphics[width=0.5\textwidth]{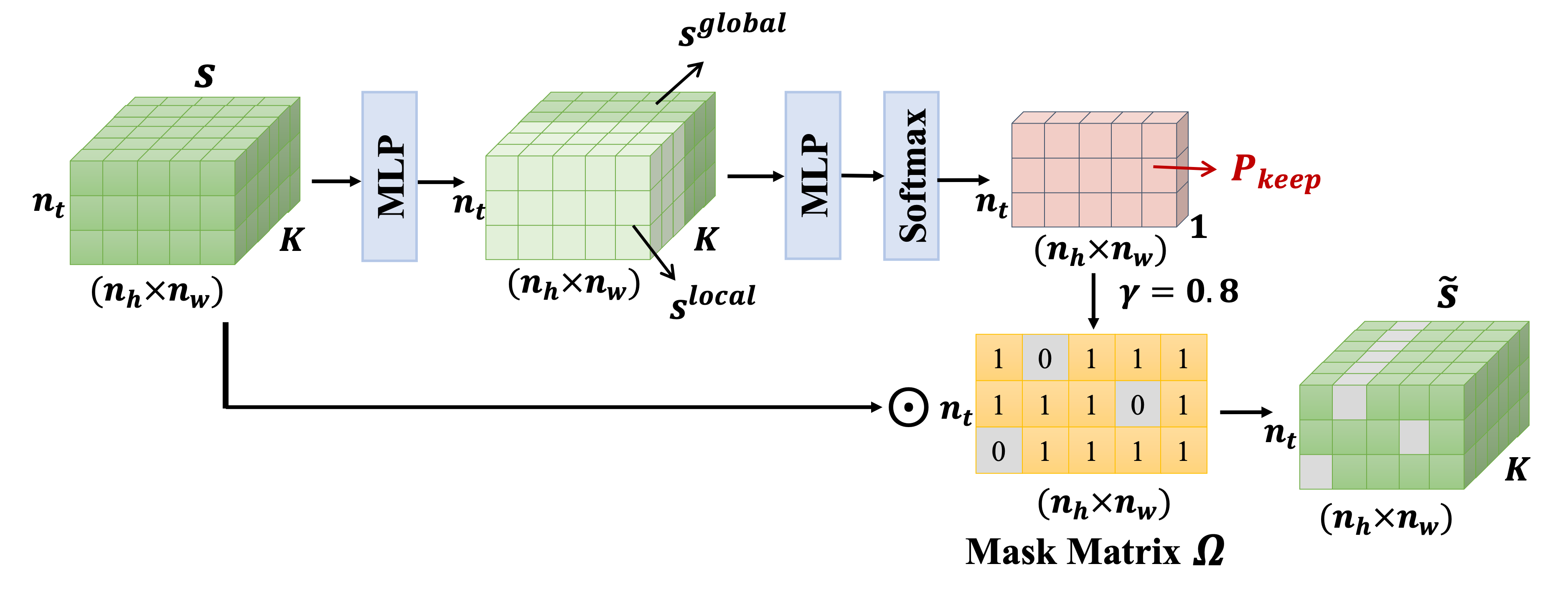}}
\caption{The structure of dynamic token selection module.}
\label{fig3}
\end{figure}

\subsection{Dynamic Token Selection Module}
In long-term video sequences, the static background in each frame remains largely consistent, leading to substantial redundancy during transmission. Consequently, this background can be regarded as content with lower semantic significance. In contrast, it is essential to prioritize the capture of motion information within the video, as this represents content with higher semantic importance. A token selection network can be developed to dynamically identify and prioritize the transmission of semantically significant content, optimizing transmission resources. 

In VDJSCC, we design a dynamic token selection module to reduce the redundancy, as illustrated in Fig. \ref{fig3}.
This module dynamically generates a mask matrix $\Omega$ based on the input video content and adjusts the encoding length according to the token keep ratio $\gamma$, enabling content-adaptive variable-length coding. To enhance the capability of the mask decisions, we take into account the influence of both local features and global features. The local features contains the information of a certain token while the global feature contains the context of the whole video sequence.
We first utilize an MLP to divide the features $\mathbf{s}$ into the local features $\mathbf{s}^{local} \in \mathbb{R}^{n_{t}\times (n_{h}\times n_{w})\times{K^{\prime } }} $ and global features $\mathbf{s}^{global}\in \mathbb{R}^{n_{t}\times 1\times{K^{\prime } }}  $. The process can be formulated as: 
\begin{equation}
    \mathbf{s}^{local} =\mathrm {MLP}(\mathbf{s}) , 
\end{equation}

\begin{equation}
    \mathbf{s}^{global} =Agg(\mathrm {MLP}(\mathbf{s})) ,
\end{equation}
where $Agg$ is the function which aggregates all tokens and can be simply implemented as the average pooling. $K^{\prime }$ is the dimension associated with token splitting. In this paper, we set $K^{\prime }$ as $K/2$.
\par Subsequently, we combine local features $\mathbf{s}^{local}$ and global features $\mathbf{s}^{global}$ to obtain the concatenated features $\mathbf{S}\in \mathbb{R}^{n_{t}\times (n_{h}\times n_{w})\times {K} }$, 
\begin{equation}
    \mathbf{S}=\left [\mathbf{s}^{local}, \mathbf{s}^{global}\right ].
\end{equation}

\noindent Then we feed $\mathbf{S}$ to another MLP to predict the probability to keep the tokens $P_{keep}\in \mathbb{R}^{ n_{t}\times (n_{h}\times n_{w}) \times {1} }$:

\begin{equation}
   P_{keep}=\mathrm {SoftMax}(\mathrm {MLP}(\mathbf{S})).
\end{equation}

\par Ultimately, we calculate the mask matrix $\Omega$ using ${P_{keep}}$ and $\gamma$ to select more semantically significant tokens from $\mathbf{S}$ for transmission. The process is written as:
\begin{equation}
\Omega=\mathbf{S}\odot \mathbb{I}(P_{keep}>1-\gamma),
\end{equation}
where $\odot$ is elementwise production, and $\mathbb{I}$ is indicator function. 
Specifically, $\Omega$ is clearly determined based on the input video content, with “0” representing the discarded features of lower semantic importance, and “1” indicating the retained features of higher semantic importance.

Moreover, the token keep ratio $\gamma$ can be adjusted to control the number of retained features, thereby enabling dynamic control over the encoding rate. Lower $\gamma$ results in lower coding rate since more tokens are discarded. At the receiver, zero padding is employed to restore the tokens to their original dimensions, resulting in the generation of feature $\mathbf{\hat{s}}$. 

\section{Experiments}
In this section, we first introduce the experiments settings, and then present the performance of the proposed VDJSCC scheme.

\subsection{Experimental Setup}\label{AA}
\par \noindent \textbf{1) Datasets} 
\par The VDJSCC model is trained and evaluated on the UCF101 dataset \cite{soomro2012dataset}, which consists of 13320 video clips across 101 action classes. All clips have fixed frame rate and resolution of 25fps and $320\times 240$ respectively. In this paper, we use the first train/test list to split the dataset into training dataset and test dataset. For each video frame, we first resize the frame to $300\times 225$, and then randomly crop into $224\times 224$. Additionally, we randomly select $16$ consecutive frames in a video to serve as the input for VDJSCC model.

\par  \noindent\textbf{2) Training Details}
\par In all experiments, the image patch size ($h$ and $w$) is set to 16, and the frame patch size ($t$) is set to 2, the channel dimension $K$ is set to 768. The spatial-temporal Transformer encoder, with a depth of $L=5$ and 12 heads, is used to extract high-dimensional features via multi-head self-attention. In the dynamic token selection module, the token keep ratio $\gamma$ is set to 0.8 during the training process. Furthermore, we train the model at different channel SNRs and evaluate each model at the same SNR, with SNR is sampled uniformly from $[1,4,7,10,13]$ dB. For each model, we use the Adam optimizer \cite{kinga2015method} with a learning rate of $10^{-4}$. The batch size is set to 4, and it takes about 1.5 week to train the model on the single RTX 3090 GPU.  

\par  \noindent\textbf{3) Evaluation Metrics}
\par In this paper, we qualify the end-to-end video transmission performance of the proposed VDJSCC model using pixel-wise metric PSNR and the perceptual metric MS-SSIM \cite{wang2003multiscale}. MS-SSIM evaluates the model at multiple scales, providing more comprehensive similarity information. 

 \begin{figure}
\centering
\subfigure[PSNR]{
\includegraphics[width=0.43\textwidth]{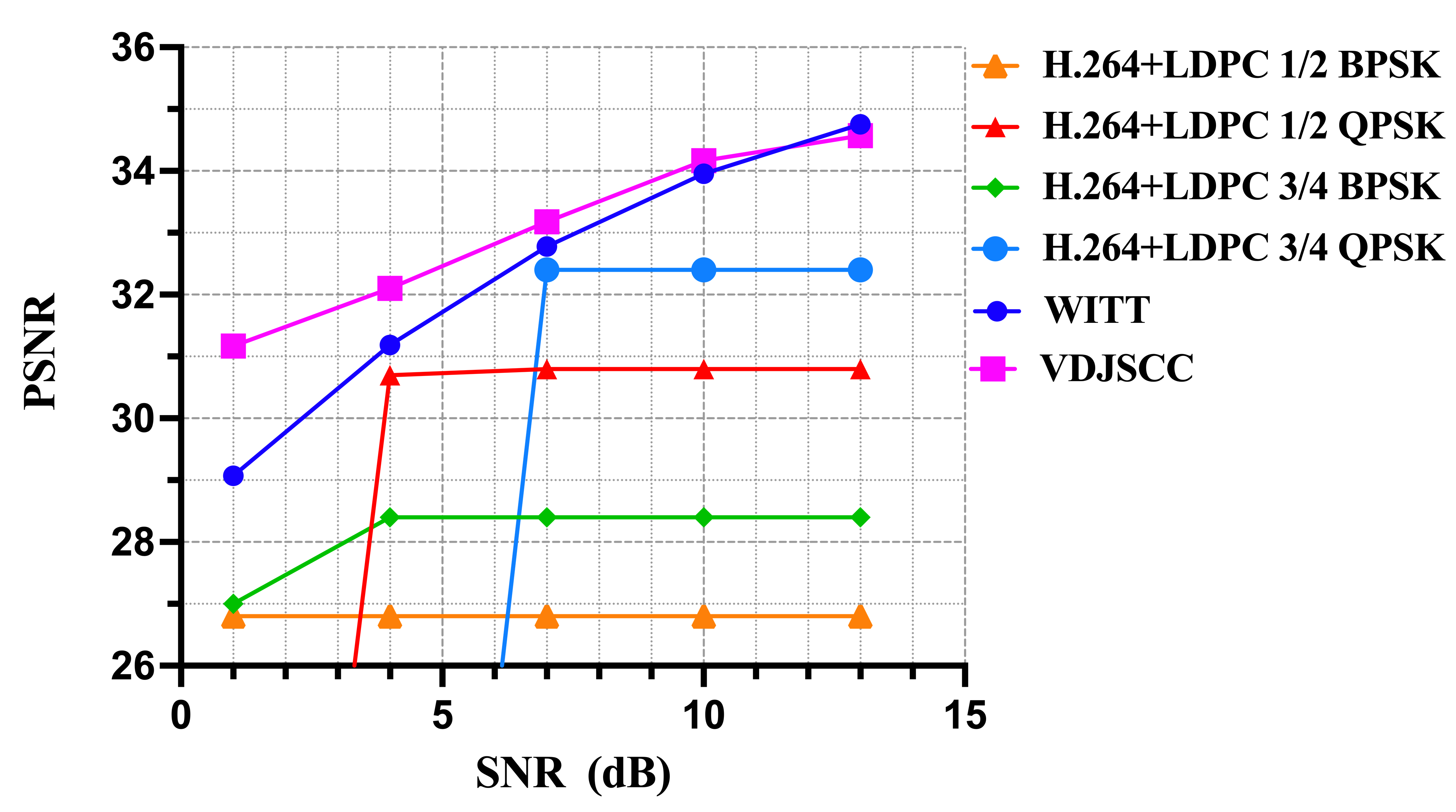}}
\subfigure[MS-SSIM]{
\includegraphics[width=0.43\textwidth]{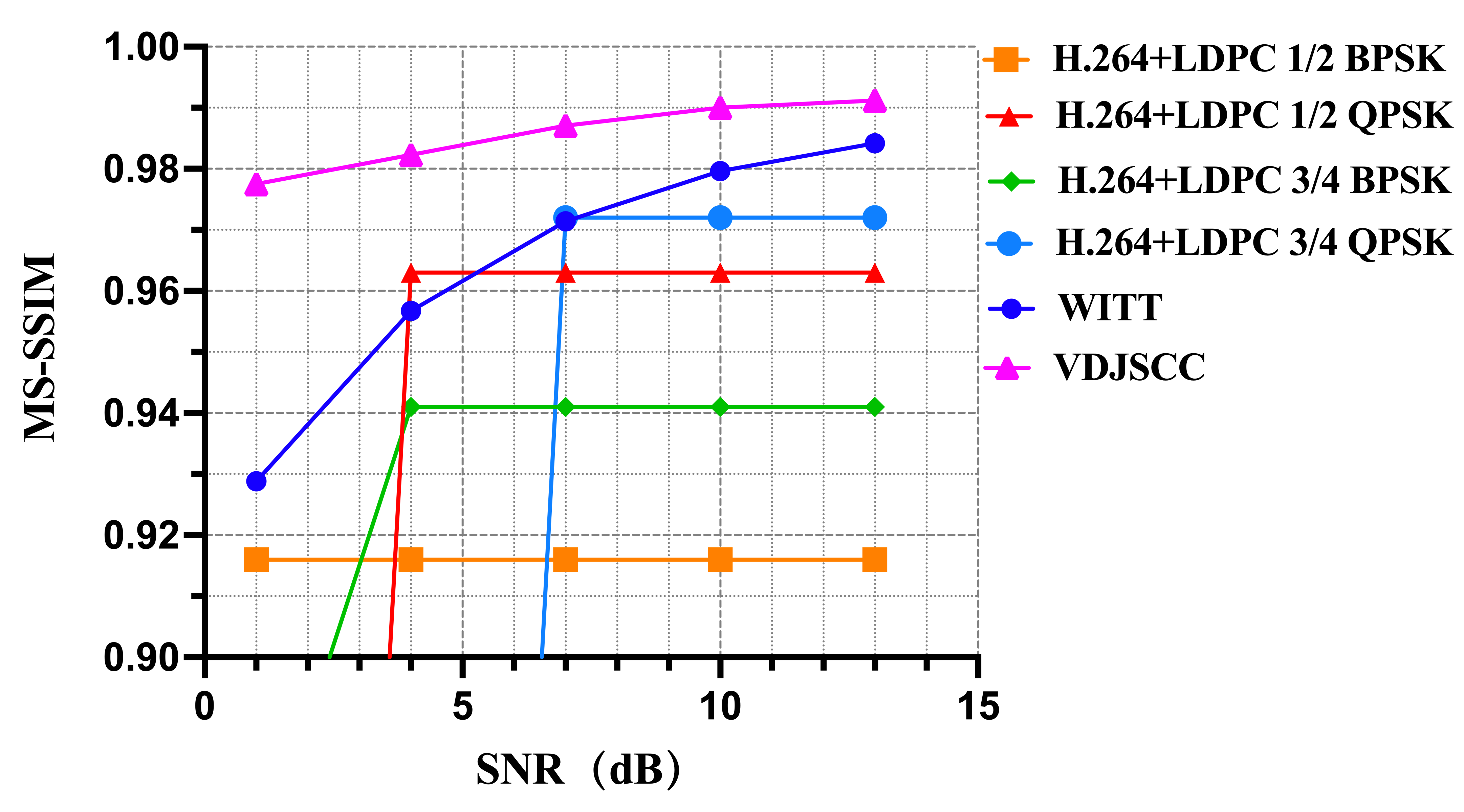}}
\caption{Performance comparison of VDJSCC to other schemes over AWGN channel cases  (CBR=0.031).}
\label{fig:example}
\end{figure}

\begin{figure}[t]
\centering
{\includegraphics[width=0.5\textwidth]{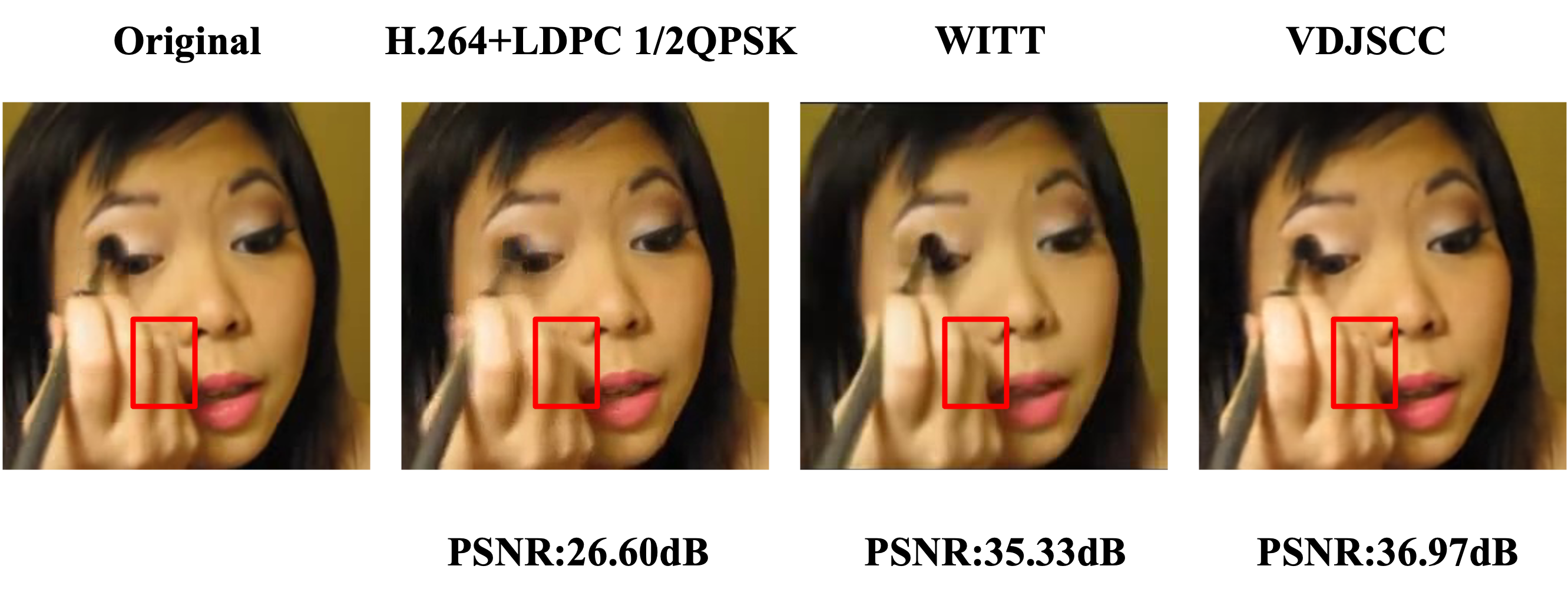}}
\caption{The examples of visual comparison under AWGN channel at SNR=13dB.}
\label{fig}
\vspace{-0.3cm}
\end{figure}

\begin{figure*}[t]
\centering
{\includegraphics[width=0.85\textwidth]{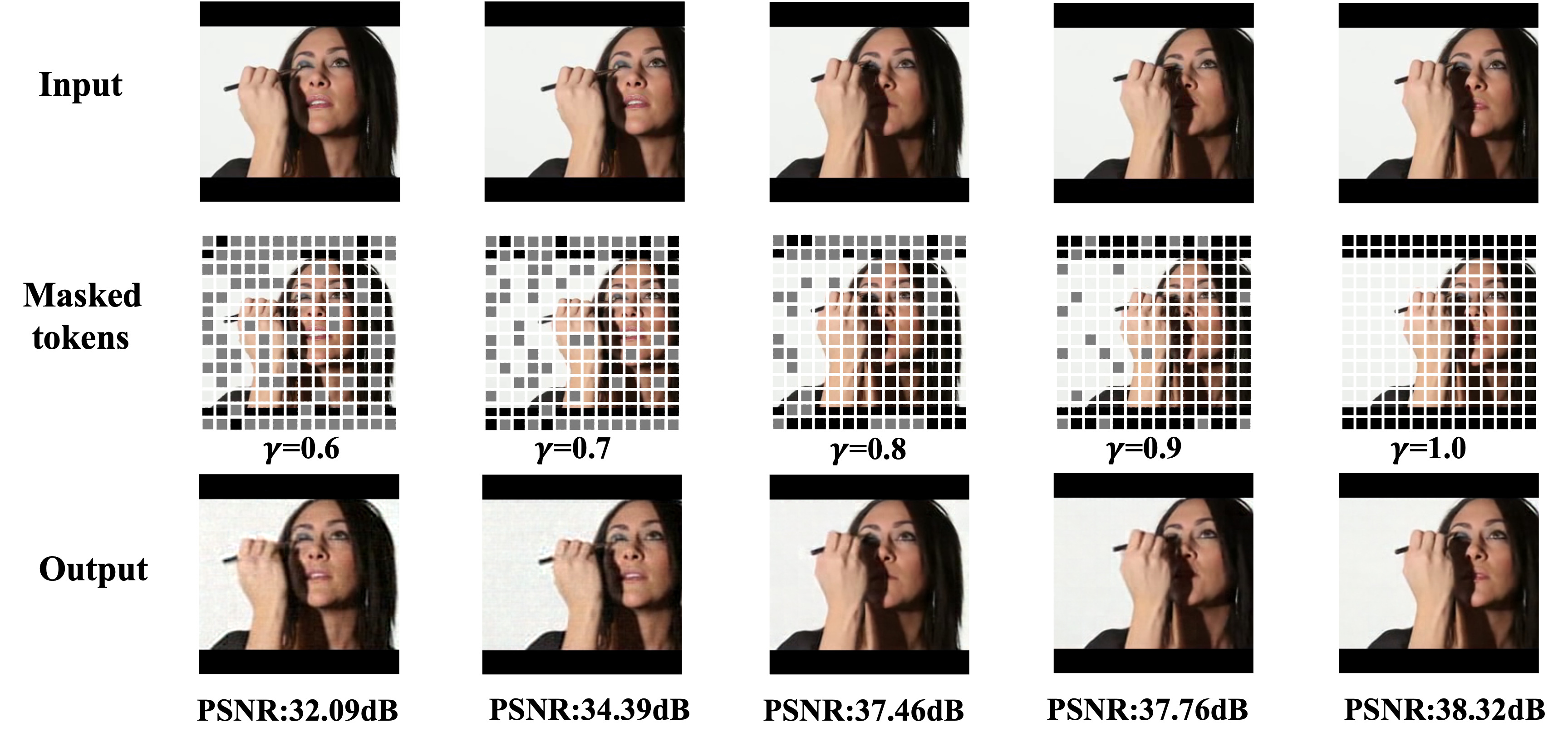}}
\caption{Visualization of token selection under AWGN channel at SNR=13dB. }
\label{fig}
\end{figure*}

\begin{figure}[t]
\centering
{\includegraphics[width=0.5\textwidth]{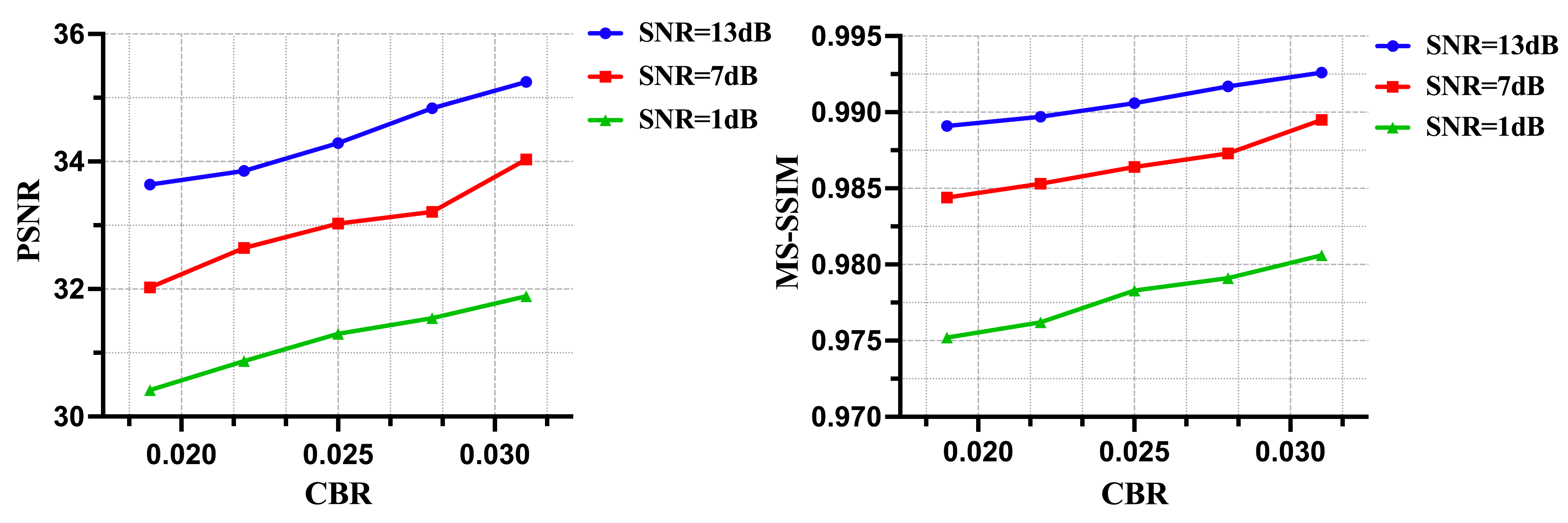}}
\caption{The performance of different channel bandwidth ratio (CBR) over AWGN channel at SNR=13dB, SNR=7dB, and SNR=1dB.}
\label{fig}
\end{figure}

\par  \noindent\textbf{4) Comparison Schemes}
\par We compare the performance of proposed VDJSCC model with the classic video coded transmission schems and a deep-learning based JSCC model. Specifically, we utilize the standard video codecs H.264 \cite{1218189} for source coding and practial LDPC \cite{richardson2018design} for channel coding. FFmpeg \cite{tomar2006converting} is used to achieve H.264 encoding.
\par In addition, we also compare the model with a state-of-the-art DeepJSCC model named WITT \cite{10094735}. WITT uses Swin Transformers as the backbone to extract the long-range information, which shows the great performance in the image reconstruction tasks. In our comparison with WITT, we first downsample the video at intervals of 10 frames and save as images, and then input the all images into the WITT model.

\subsection{Experimental Results}
\par  \noindent\textbf{1) Analysis of Video Reconstruction Performance Under Different SNR.}

Fig. 4 shows the PSNR and MS-SSIM performance versus SNR over the AWGN channel.  For the H.264+LDPC, we compare the impact of different modulation methods and code rates. The proposed VDJSCC model significantly outperforms classical video transmission schemes, avoiding the \textit{cliff-effect}. When the SNR is 1dB, VDJSCC achieves a PSNR of 31.17dB, compared to WITT's 29.07dB, indicating a notable improvement in reconstruction quality. However, WITT slightly surpasses VDJSCC in PSNR when the SNR exceeds 10dB due to its focus on individual frame processing, which neglects temporal correlations and increases computational resources.
In terms of MS-SSIM, VDJSCC consistently outperforms other schemes, particularly at low SNR levels, demonstrating superior perceptual quality with higher visual similarity.
\par Fig. 5 vividly demonstrates the visual comparison of VDJSCC and the classical video transmission scheme under AWGN channel at SNR=13dB. The proposed VDJSCC preserves clearer details and achieves higher reconstruction quality.

\par  \noindent\textbf{2) Analysis of Video Compression Performance.}
\par Next, we explore the impact of CBR on the performance of the proposed VDJSCC scheme. In this part, we set different token keep ratio to alter the CBR, adjusting the CBR to 0.031 when the token keep ratio $\gamma$=1.0. Fig. 6 provides the visual results of token selection of the same input video under AWGN channel at SNR=13dB. The model dynamically generates the mask matrix based on the input videos and token keep ratio, effectively masking less semantically important background information while retaining key portrait details. These results demonstrate the model's ability to achieve high-quality video reconstruction.

\par Furthermore, in Fig. 7, we present the performance of PSNR and MS-SSIM of different CBR. The models are trained under the AWGN channel at SNR=13dB, SNR=7dB, and SNR=1dB. It is shown that, the PSNR and MS-SSIM increase along with the increase of CBR. Specifically, with a CBR of 0.019, the PSNR exceeds 30dB, indicating good video reconstruction quality even with nearly half of the tokens masked. Notably, token selection allows for dynamic adjustment of coding length based on the token keep ratio, enabling variable-length encoding and efficient bandwidth utilization.

\par  \noindent\textbf{3) Ablation Study and Computational Cost Comparison.}

\par Last but not least, we present the results of ablation study to evaluate the influence of different modules. Table {\Rmnum{1}} illustrates the performance of base models, all trained under an AWGN channel at SNR=13dB, with CBR=0.031. For models with the token selection module, the token keep ratio $\gamma$ is set to 0.8. "VDJSCC w/o multi-scale" refers to the model without patch merging, where video frames are transmitted directly into the 10-layer spatial-temporal Transformer encoder. "VDJSCC w/o token selection" means the model without dynamic token selection, where all tokens are transmitted. Comparative results clearly show that the proposed multi-scale method can improve the reconstruction quality by 3.52dB. However, the token selection module slightly reduces PSNR by 0.62dB, an acceptable trade-off given its bandwidth-saving benefits.
\par With respect to complexity, the average inference time of the proposed VDJSCC is approximately 77.42ms using a RTX 3090 GPU. 
 By comparison, the classical H.264 scheme runs at speeds ranging from 1fps to over 100fps, depending on coding settings, with typical encoding rates between 10-20fps. 
Although the multi-scale method requires more computational resources, it results in an approximately $10\%$ increase in PSNR.
The ablation study indicates that the proposed VDJSCC method achieves high-quality video reconstruction while optimizing bandwidth and computational resources.

\begin{table}[htb]
\centering
\caption{Results on ablation study. The base models are all trained under AWGN channel at SNR=13dB, CBR=0.031, and $\gamma$ =0.8.}
\label{tab:my-table}
\renewcommand\arraystretch{2}
\resizebox{\columnwidth}{!}{%
\begin{tabular}{ccccc}
\hline
\rowcolor[HTML]{ECF4FF} 
\textbf{Base Model}              & \textbf{PSNR} & \textbf{MS-SSIM} & \textbf{Inference time (ms)} & \textbf{FLOPS (G)}\\ \hline
VDJSCC                  & 34.54                      & 0.9906  & 77.42   &321.8                    \\
w/o multi-scale                  & 31.02 {\color[HTML]{3531FF} (-3.52)}  & 0.9796 {\color[HTML]{3531FF}(-0.0110)}  & 60.05  & 252.7                       \\
w/o token selection              & 35.16 {\color[HTML]{3531FF}(+0.62)}  & 0.9923 {\color[HTML]{3531FF}(+0.0017)}  & 78.09   & 320.2                       \\
w/o multi-scale\&token selection & 32.42 {\color[HTML]{3531FF}(-2.12)}  & 0.9843 {\color[HTML]{3531FF}(-0.0063)}  & 61.89   & 251.2                      \\ \hline
\end{tabular}%
}
\end{table}

\section{Conclusion}
This paper has proposed a novel DeepJSCC method to achieve end-to-end wireless videos transmission. The proposed VDJSCC scheme utilizes the multi-scale spatial-temporal Transformer encoder to explore multi-scale semantic information and obtain a new spatial-temporal representation. Moreover, a dynamic token selection algorithm has been employed to mask tokens with less semantic importance, facilitating content-adaptive variable-length encoding. 
Extensive experimental results have shown that the proposed VDJSCC scheme outperforms the classical separation-based video transmission scheme H.264+LDPC by a large margin and overcomes the \textit{cliff-effect}. We have also proved that the multi-scale method greatly enhances video reconstruction quality. Furthermore, the token selection module has saved certain bandwidth resources and has accelerated the model's inference time while ensuring the reconstruction quality. 

\section*{Acknowledgment}
This work is supported in part by the National Natural Science Foundation of China (NSFC) under Grant 62101307, U23B2052, 62071061 and in part by the Fundamental Research Funds for the Central Universities 2023RC78.

\small
\bibliographystyle{IEEEtran}
\bibliography{ref}


\end{document}